\documentstyle[prd,eqsecnum,preprint,tighten,aps]{revtex}
\begin{document}
\draft
\title{Casimir energy of a dilute
dielectric ball in the mode summation method}
\author{G.~Lambiase\thanks{Electronic address: lambiase@sa.infn.it},
G. Scarpetta\thanks{Electronic address: scarpetta@sa.infn.it}}
\address{Dipartimento di Scienze Fisiche ``E.R. Caianiello'',
 Universit\'a di Salerno, 84081 \\ Baronissi (SA), Italy.\\
 INFN, Sezione di Napoli--Napoli, Italy.\\
 International Institute for Advanced Scientific Studies--Vietri sul
 Mare, Italy.}
\author{V.V.~Nesterenko\thanks{Electronic address: nestr@thsun1.jinr.ru}}
\address{Bogoliubov Laboratory of Theoretical Physics,
Joint Institute for Nuclear Research, 141980 Dubna,  Russia}
\date{\today}
\maketitle
\begin{abstract}
In the  $(\varepsilon_1-\varepsilon_2)^2$--approximation the Casimir
energy of a dilute dielectric ball is derived using a simple and
clear method of the mode summation. The addition theorem for the
Bessel functions enables one to present in a closed form the sum over
the angular momentum before the integration over the imaginary
frequencies. The linear  in $(\varepsilon_1-\varepsilon_2)$
contribution into the vacuum energy is removed by an appropriate
subtraction.  The role of the contact terms used in other approaches
to this problem is elucidated.
\end{abstract}

\section{Introduction}
The progress in calculation of the Casimir energy is rather slow.  In
his pioneer paper \cite{Cas} in 1948 H.~B.~G.\ Casimir calculated the
vacuum electromagnetic energy for the most simple boundary
conditions, for two parallel perfectly conducting plates placed in
vacuum. Dielectric properties of the media separated by plane
boundaries do not add new mathematical difficulties \cite{Lif}.
However the first result on the calculation of the Casimir energy for
the non-flat boundaries was obtained only in 1968. By computer
calculations, lasted 3 years, T. H. Boyer found the Casimir energy of
a perfectly conducting spherical shell \cite{Boyer}. Account of
dielectric and magnetic properties of the media in calculations of
the vacuum energy for nonflat interface leads to new principal
difficulties or, more precisely, to a new structure of divergencies.

     The calculation of the Casimir energy in a special case, when
both the material media have the same velocity of light, proves to
be, from the mathematical stand point, exactly the same as for
perfectly conducting shells placed in vacuum and having the shape of
the interface between these media
\cite{BNP,NP-cyl,MNN,BC,BC1,BC2,BC3,BY,Klich,Klich-2,Klich-FMR}.

     The first attempt to calculate  the Casimir energy of a
dielectric compact ball has been undertaken by K.~A.\ Milton in 1980
\cite{Milton}.  And only just recently the vacuum electromagnetic
energy of a dilute dielectric ball was found\footnote{It is worth
noting that the first right (but rough and not rigorous) estimation
of the Casimir energy of a dilute dielectric ball has been done in
the paper \cite{BNP}.}  \cite{MNg2,Barton,BMM,BM,Mar}. The light
velocity is discontinuous on the surface of such a ball.  In
Ref.~\cite{Bordag} the analysis of the divergencies, which appear in
calculation of the Casimir energy of a dielectric ball, has been
carried out  by determining the relevant heat kernel coefficients.
The role of dispersion in this problem is now under study also
\cite{BC1,BM,BS,BSS,NP-dis,Candelas}.

     Under these circumstances it is of a indubitable interest to
develop  new methods for calculating the vacuum energy for non-flat
boundaries with allowance for the material characteristics of the
media. It is this aim that is pursued in the present paper.

     Practically all the calculations of the Casimir energy for the
boundaries with spherical or cylindrical symmetries use the uniform
asymptotic expansion for the Bessel functions. In place of this we
are employing the addition theorem for these functions that enables
one to accomplish the summation over the values of the angular
momentum exactly, i.e., in a closed form. In addition, the starting
point in our calculation is a simple and clear representation of the
vacuum energy  as a half sum of the eigenfrequencies of
electromagnetic oscillations connected with a dielectric ball (a
global approach).  This fact is also important  due to the following
consideration.  From the mathematical stand point, the most
consistent method for treating the divergencies in calculations of
the  vacuum energy is the zeta regularization technique \cite{Od}. In
this approach, one proceeds from the sum of the eigenfrequencies.  In
Refs.\ \cite{BMM,BM,Mar} the Casimir energy of a dielectric ball has
been calculated by making use of the Green's function method. An
essential point there was the account of the so called contact terms.
These terms encounter the expression for the vacuum energy being
outside the logarithm function \cite{MNg2,BMM,MNg1,M}, therefore they
do not appear when one proceeds from the sum of the eigenfrequencies.

     It is worth noting that the results of the Casimir energy
calculation for a dilute dielectric ball, accomplished in the
framework of the quantum field theory \cite{BMM,BM}, coincide with
those obtained by summing up the van der Waals interactions between
individual molecules inside the ball \cite{MNg2} and by applying a
special perturbation theory, where the dielectric ball is treated as
a perturbation of the electromagnetic field in unbounded empty space
\cite{Barton}.

     The layout of the paper is as follows. In Sec.\ II the
derivation of the integral representation for the vacuum energy is
given by the mode sum and contour integration. The subtraction
procedure that gives the renormalized Casimir energy in the $\Delta
n^2$-approximation is discussed in detail as well as its physical
justification. The addition theorem for the Bessel functions enables
one to carry out the sum over the angular momentum in a closed form.
It leads to an exact (in the $\Delta n^2$-approximation) value of the
Casimir energy of a dilute dielectric ball. In the Conclusion (Sec.\
III) the method proposed here for calculating the Casimir energy is
briefly discussed, as well as the implications of the obtained
results concerning, specifically, the elucidation of the role of the
contact terms used in other approaches to this problem. In the
Appendix the analysis of the divergencies is accomplished revealing
an important relation between the linear and quadratic in $\Delta n$
contributions into the vacuum energy. It is this relation that
provides a simple and effective scheme of calculations which is
followed in this paper.

\section{Mode summation for vacuum electromagnetic energy of a
dilute dielectric ball}

     We shall consider a solid ball  of radius $a$ placed in an
unbounded uniform medium. The nonmagnetic materials making up the
ball and its surroundings are characterized by permittivity
$\varepsilon_1$ and $\varepsilon_2$, respectively. It is assumed that
the conductivity in both the media is zero.  The natural system of
units is used where $c=\hbar=1$.

     We shall proceed from the standard definition of the vacuum
energy as the sum over the eigenfrequencies of electromagnetic
oscillations \cite{PRep}
\begin{equation}\label{def}
  E=\frac{1}{2}\sum_s(\omega_s-\overline{\omega}_s)\,.
\end{equation}
Here $\omega_s$ are the classical frequencies of the electromagnetic
field for the boundary conditions described above, and the
frequencies $\bar{\omega}_s$ correspond to a certain limiting
boundary conditions that will be specified below.

     The sum $(1/2)\sum_s\overline{\omega}_s$ in Eq.\ (\ref{def})
plays the same role as the counter terms in the  standard
renormalization procedure in quantum field theory \cite{Bsh}. However
in  the renormalizable field models, considered in unbounded
Minkowski space-time, the explicit form of these counter terms is
known (at least, it is known the algorithm of their construction at
each order of perturbation theory).  Unlike this, there are no
general rules for obtaining the terms that should be subtracted when
calculating the vacuum energy. Therefore, in a new problem on
calculating the Casimir energy it is necessary to specify the
boundary conditions, determining the frequencies
$\overline{\omega}_s$, anew, appealing to some physical
considerations.

     In the case of the plane geometry of boundaries or when
considering the Casimir effect for  distinct bodies it is sufficient
to subtract in Eq.\ (\ref{def}) the contribution of the Minkowski
space~\cite{PRep,Bordag}. In the problem at hand it implies to take
the limit $a\to \infty$, i.e., that the medium 1 tends to fill the
entire space. But it turns out that this subtraction is not
sufficient because the linear in $\varepsilon_1-\varepsilon_2$
contribution into the vacuum energy retains.  Further, we assume
that the difference  $\varepsilon_1-\varepsilon_2$ is small and
content ourselves only with the
$(\varepsilon_1-\varepsilon_2)^2$-terms.

     The necessity to subtract the contributions into the vacuum
energy linear in  $\varepsilon_1-\varepsilon_2$ is justified by the
following consideration.  The Casimir energy of a dilute dielectric
ball can be thought of as the net result of the van der Waals
interactions between the molecules making up the ball \cite{MNg2}.
These interactions are proportional to the dipole momenta of the
molecules, i.e., to the quantity $(\varepsilon_1-1)^2$. Thus, when a
dilute dielectric ball is placed in the vacuum, then its Casimir
energy should be proportional to $(\varepsilon_1-1)^2$. It is natural
to assume that when such a dielectric ball is surrounded by an
infinite dielectric medium with permittivity $\varepsilon_2$, then
its Casimir energy should be proportional to
$(\varepsilon_1-\varepsilon_2)^2$. The physical content of the
contribution into the vacuum energy linear in
$\varepsilon_1-\varepsilon_2$ has been investigated in the framework
of the microscopic model of the dielectric media (see Ref.\
\cite{MSS} and references therein). It has been shown that this term
represents the self-energy of the electromagnetic field attached to
the polarizable particles or, in more detail, it is just the sum of
the individual atomic Lamb shifts. Certainly this term in the vacuum
energy should be disregarded when calculating the Casimir energy
which is originated in the electromagnetic interaction between
different polarizable particles or atoms
\cite{Barton,BMM,Barton-dis,H-Brevik,H-Brevik-A}.

     Further, we put for sake of symmetry
\begin{equation}\label{dn}
  \sqrt{\varepsilon_1}=n_1=1+\frac{\Delta n}{2}\,,\quad
 \sqrt{\varepsilon_2}=n_2=1-\frac{\Delta n}{2}\,.
\end{equation}
Here $n_1$ and $n_2$ are the refractive indices of the ball and of
its surroundings, respectively, and it is assumed that $\Delta n<<1$.
From here it follows, in particular, that
\begin{equation}\label{epn}
  \varepsilon_1-\varepsilon_2=(n_1+n_2)(n_1-n_2)=2\Delta n\,.
\end{equation}
Thus, using the definition (\ref{def}) we shall keep in mind that
really two subtractions should be done: first the contribution,
obtained in the limit $a\to \infty$, has to be subtracted and then
all the terms linear in $\Delta n$ should also be removed.

We present the vacuum energy defined by Eq.\ (\ref{def}) in terms of
the contour integral in the complex frequency plane. The details of
this procedure can be found in Refs. \cite{BNP,LNBor,new1,new2}. Upon
the contour deformation one gets
\begin{equation}\label{integral}
  E=-\frac{1}{2\pi}\sum_{l=1}^{\infty}(2l+1)\int_0^{K}dy\,y\,\frac{d}{dy}
  \ln\frac{\Delta_l^{\text{TE}}(iay)\Delta_l^{\text{TM}}(iay)}
  {\Delta_l^{\text{TE}}(i\infty)\Delta_l^{\text{TM}}(i\infty)}\,,
\end{equation}
where $\Delta_l^{\text{TE}}(iay)$ and $\Delta_l^{\text{TM}}(iay)$
are the left-hand sides of the equations determining the
frequencies of the electromagnetic field
\begin{equation}\label{freq}
  \Delta_l^{\text{TE}}(a\omega)=0\,,\quad
  \Delta_l^{\text{TM}}(a\omega)=0\,.
\end{equation}
For pure imaginary values of the frequency variable $\omega=iy$
(these values are needed in Eq.\  (\ref{integral})), the expressions
$\Delta_l^{\text{TE}}(iay)$ and $\Delta_l^{\text{TM}}(iay)$ are
defined by
\begin{eqnarray}\nonumber
  \Delta_l^{\text{TE}}(iay) &=&
  \sqrt{\varepsilon_1}s_l^{\prime}(k_1a)e_l(k_2a)-
  \sqrt{\varepsilon_2}s_l(k_1a)e_l^{\prime}(k_2a)\,, \\
  \Delta_l^{\text{TM}}(iay) &=&
  \sqrt{\varepsilon_2}s_l^{\prime}(k_1a)e_l(k_2a)-
  \sqrt{\varepsilon_1}s_l(k_1a)e_l^{\prime}(k_2a)\,,\label{TETM}
\end{eqnarray}
where $k_i=\sqrt{\varepsilon_i}\,y$, $i=1,2$, and $s_l(x)$,
$e_l(x)$ are the modified Riccati--Bessel functions \cite{AS}
\begin{equation}\label{RB}
  s_l(x)=\sqrt{\frac{\pi x}{2}}\,I_{\nu}(x)\,,\quad
  e_l(x)=\sqrt{\frac{2 x}{\pi}}\,K_{\nu}(x)\,,\quad
  \nu=l+\frac{1}{2}\,.
\end{equation}
The prime in Eq.\ (\ref{TETM}) stands for the differentiation with
respect to the argument of the Riccati--Bessel functions.

In Eq.\ (\ref{integral}) we have introduced cutoff $K$  in
integration over the frequencies. This regularization is natural
in the Casimir problem because physically it is clear that the
photons  of a very short wave length do not contribute into the
vacuum energy since they do not ``feel'' the boundary between the
media with different permittivities $\varepsilon _1$ and
$\varepsilon _2$. In the final expression the regularization
parameter $K$ should be put to tend to infinity, the divergencies,
that may appear here, being  canceled by appropriate counter
terms.

The numerator (denominator) in the logarithm function in Eq.\
(\ref{integral}) is responsible for the first (second) term in the
initial formula (\ref{def}). For brevity we write in Eq.\
(\ref{integral}) simply $\Delta_l(i\infty)$ instead of
$\lim_{a\to\infty}\Delta_l(iay)$. Taking into account the
asymptotics of the Riccati--Bessel functions
\[
  s_l(x)\simeq \frac{1}{2}\,e^x\,, \quad e_l(x)\simeq e^{-x}\,,
  \quad x\to \infty\,,
\]
we obtain
\begin{equation}\label{asymp2}
 \Delta_l^{\text{TE}}(i\infty)\Delta_l^{\text{TM}}(i\infty)=
 -\frac{(n_1+n_2)^2}{4}\,e^{2(n_1-n_2)y}\,.
\end{equation}
Upon substituting Eqs.\ (\ref{TETM}) and (\ref{asymp2}) into Eq.\
(\ref{integral}) and changing the integration variable $ay\to y$,
we cast Eq.\  (\ref{integral}) into the form (see Eq.\  (tef{E2}) in Ref.
\cite{BNP})
\begin{eqnarray}\label{final}
 E&=&-\frac{1}{2\pi a}\sum_{l=1}^{y_0}(2l+1)\int_0^{y_0}dy
 \,y\,\frac{d}{dy}
  \ln\left\{ \frac{4e^{-2(n_1-n_2)y}}{(n_1+n_2)^2} \right. \\
  & & \left. [n_1n_2((s_l^{\prime}e_l)^2+
  (s_le_l^{\prime})^2)-(n_1^2+n_2^2)s_ls_l^{\prime}e_le_l^{\prime}]\right\}\,,
  \nonumber
\end{eqnarray}
where $s_l\equiv s_l(n_1y)$, $e_l\equiv e_l(n_2y)$, $y_0=a K$.

It should be noted here that in Eq.\ (\ref{final}) only the first
subtraction is accomplished, which removes the contribution into
the vacuum energy obtained when $a\to\infty$. As noted above, for
obtaining the final result all the terms linear in $\Delta n$
should be discarded also.

Further it is convenient to rewrite Eq.\  (\ref{final}) in the form
\begin{equation}\label{E1E2}
 E=E_1+E_2
\end{equation}
with
\begin{eqnarray}\label{E1}
 E_1&=&\frac{\Delta n}{2\pi
 a}\sum_{l=1}^{\infty}(2l+1)\int_0^{y_0}y\,dy\,, \\
 E_2&=&-\frac{1}{2\pi a}\sum_{l=1}^{\infty}(2l+1)\int_0^{y_0}dy\,
 y\,\frac{d}{dy}\ln\left[W_l^2(n_1y, n_2y)-\frac{\Delta n^2}{4}\,
 P_l^2(n_1y, n_2y)\right]\,,
 \label{E2}
\end{eqnarray}
where
\begin{eqnarray}
  W_l(n_1y, n_2y)&=&s_l(n_1y)e_l^{\prime}(n_2y)-s_l^{\prime}(n_1y)e_l(n_2y)\,,
  \label{W} \\
  P_l(n_1y, n_2y)&=&s_l(n_1y)e_l^{\prime}(n_2y)+s_l^{\prime}(n_1y)e_l(n_2y)\,.
  \label{P}
\end{eqnarray}
The term $E_1$ accounts for only the expression $\exp (-2\Delta n
\,\,y)$ in the argument of the logarithm function in Eq.\
(\ref{final}) and it appears as a result of subtracting the Minkowski
space contribution into the Casimir energy (the sum with $\bar
\omega_s$ in Eq.\ (\ref{def}) and the denominator in Eq.\
(\ref{integral})).

It is worth noting that the term $E_1$ is exactly  the
Casimir energy
considered by Schwinger in his attempt to explain the sonoluminescence
\cite{Schw}.
Really, introducing the cutoff $K$ for frequency integration and the cutoff
$y=\omega/a$ for the angular momentum summation we arrive at the result
\begin{equation}\label{Schwinger}
  E_1=\frac{\Delta n}{\pi a}\int_0^{aK}y\,dy\, \sum_{l=1}^{\infty}
  \left(l+\frac{1}{2}\right)\sim \frac{\Delta n}{2\pi a}\int_0^{aK}y^3\,dy=
  \Delta n\,\frac{K^4a^3}{8\pi}\,.
\end{equation}
We have substituted here the summation over $l$ by integration. Up to
the multiplier
$(-2/3)$ it is exactly the Schwinger value for the Casimir energy of a
ball ($\varepsilon_1=1$) in water ($\sqrt{\varepsilon_2}\simeq 4/3$)
\cite{MNg1}.  The term linear in $\Delta n$ and of the same structure
was also derived in Refs.\ \cite{Barton,Barton-dis,H-Brevik}.
As it was explained above the energy $E_1$  should be discarded.

In our calculation, we content ourselves with the $\Delta
n^2$-approximation. Hence, in Eq.\  (\ref{E2}) one can put
$P^2_l(n_1y, n_2y)\simeq P^2_l(y,y)$ and keep in expansion of the
logarithm function only the terms proportional to $\Delta n^2$. In
this approximation, the contributions of $W_l^2$ and $P_l^2$ into
the vacuum energy are additive
\begin{equation}\label{EWP}
  E^{\text{ren}}=E_W^{\text{ren}}+E_P^{\text{ren}}\,.
\end{equation}
In the Appendix it is shown that for obtaining the $\Delta
n^2$--contribution into the Casimir energy of the function $W_l^2$ in the
argument of the logarithm in Eq.\  (\ref{E2}), it is sufficient to
calculate the $\Delta n^2$--contribution of the function $W_l^2$
alone but changing the sign of this contribution to the opposite one
(see Eq.\ (\ref{A20})).
Hence,
\begin{equation}\label{EW} E_W=\frac{1}{2\pi
  a}\sum_{l=1}^{\infty}(2l+1)\int_0^{y_0}dy \,y\,\frac{d}{dy}
 W_l^2(n_1y, n_2y)\,,
\end{equation}
and only the $\Delta n^2$-term being preserved in this expression.

For $E_P$ we have
\begin{equation}\label{EP}
  E_P=\frac{\Delta n^2}{8\pi
  a}\sum_{l=1}^{\infty}(2l+1)\int_0^{y_0}\,dy\,
 y\frac{d}{dy} P_l^2(n_1y, n_2y)\,.
\end{equation}
Usually, when calculating the vacuum energy in the problem with
spherical symmetry, the uniform asymptotic expansion of the Bessel
functions is used \cite{AS}. As a result, an approximate value of
the Casimir energy can be derived, the accuracy of which depends
on the number of terms preserved in the asymptotic expansion.

We shall persist in another way employing the technique
of the paper~\cite{Klich}. By making use of the addition theorem
for the Bessel functions \cite{AS}, we first do the
summation over the angular momentum $l$ in Eq.\  (\ref{E2}) and only
after that we will integrate over the imaginary frequency~$y$. As a
result, we obtain an exact (in the $\Delta n^2$--approximation) value
of the Casimir energy in the problem involved.

Further  the following  addition theorem for the Bessel
functions \cite{AS} will be used
\begin{equation}\label{addition}
  \sum_{l=0}^{\infty}(2l+1)s_l(\lambda
  r)e_l(\lambda\rho)P_l(\cos\theta)=\frac{\lambda r\rho}{R}\,
e^{-\lambda R}
  \equiv {\cal D}\,,
\end{equation}
where
\begin{equation}\label{R}
  R=\sqrt{r^2+\rho^2-2r\rho\cos\theta}\,.
\end{equation}
Differentiating the both sides of Eq.\  (\ref{addition}) with
respect to $\lambda r$ and squaring the result we deduce
\begin{equation}\label{D}
  \sum_{l=0}^{\infty}(2l+1)[s_l^{\prime}(\lambda
  r)e_l(\lambda\rho)]^2=\frac{1}{2r \rho}\int_{r-\rho}^{r+\rho}
  \left(\frac{1}{\lambda}\,\frac{\partial{\cal D}}{\partial r}
 \right)^2R\,dR\,.
\end{equation}
Here the orthogonality relation for the Legendre polynomials
\[
  \int_{-1}^{+1}P_l(x)P_m(x)dx=\frac{2\delta_{lm}}{2l+1}
\]
has been taken into account. Now we put
\begin{equation}\label{ndel}
  \lambda=y\,,\quad r=n_1=1+\frac{\Delta n}{2}\,,\quad
  \rho=n_2=1-\frac{\Delta n}{2}\,.
\end{equation}
Applying Eq.\ (\ref{D}) and analogous ones, we
derive \begin{eqnarray}\nonumber
  \sum_{l=1}^{\infty}(2l+1)W_l^2(n_1y, n_2y)&=&
  \frac{1}{2r\rho\lambda^2}\int_{r-\rho}^{r+\rho}R\,dR\left({\cal D}_r-
  {\cal D}_{\rho}\right)^2-e^{2\Delta n y} \\
     &=& \frac{\Delta n^2}{8}\int_{\Delta n}^{2}\frac{e^{-2yR}}{R^5}
     \left(4+R^2+4yR-yR^3\right)^2dR-e^{2\Delta n y}\,, \label{WD} \\
  \sum_{l=1}^{\infty}(2l+1)P_l^2(y,
  y)&=&\frac{1}{2}\int_0^2\left[\frac{\partial}{\partial y}
  \left(\frac{y}{R}\,e^{-yR}\right)\right]^2R\,dR-e^{-4y}\,. \label{PD}
\end{eqnarray}
Here ${\cal D}_r$ and ${\cal D}_{\rho}$ stand for the results of
the partial differentiation of the function ${\cal D}$ in Eq.\
(\ref{addition}) with respect to the corresponding variables and with
the subsequent substitution of (\ref{ndel}). The last terms in Eqs.\
(\ref{WD}) and (\ref{PD}) are $W_0^2(n_1y, n_2y)$ and $P_0^2(y,
y)$, respectively. As it was stipulated before, in Eq.\ (\ref{WD})
we have to keep
only the terms proportional to $\Delta n^2$ and in
Eq.\  (\ref{PD}) we have put $\Delta n=0$.

The calculation of the contribution $E_P$ to the Casimir energy is
straightforward.  Upon differentiation of the right-hand side of
Eq.\ (\ref{PD}) with respect to $y$, the integration over $dR$ can
be done here. Substitution of this result into Eq.\ (\ref{EP})
gives
\begin{equation}\label{EP1}
  E_P=-\frac{\Delta n^2}{2\pi
  a}\left(-\frac{1}{4}\right)\int_0^{y_0}dy\,\left[e^{-4y}\left(2y^2+2y+
  \frac{1}{2}\right)-\frac{1}{2}\right]\,.
\end{equation}
The term $(-1/2)$ in the square brackets in Eq.\  (\ref{EP1}) gives
rise to the divergence\footnote{This divergence has the same origin
as those arising in summation over $l$ when the uniform asymptotic
expansions of the Bessel functions are used \cite{BMM,BM}. The
technique employed here is close to the multiple scattering expansion
\cite{BD}, where these divergencies are also subtracted.} when  $y_0
\to \infty$
\begin{equation}
\label{ct1}
E_P^{\text{div}} = - \frac{\Delta n^2}{16\pi a}y_0\,{.}
\end{equation}
Therefore we have to subtract it with the result
\begin{equation}\label{EPF}
  E_P^{\text{ren}}=E_P-E_P^{\text{div}}
=\frac{5}{128}\,\frac{\Delta n^2}{\pi a}\,.
\end{equation}
As far as the expression (\ref{WD}), it is convenient to
substitute it into Eq.\  (\ref{EW}), to do the integration over $y$
and only after that to address the integration over $dR$
\begin{eqnarray}
 \lefteqn{\frac{\Delta n^2}{8}\int_{\Delta n}^2 dR \int_0^{\infty}
  dy\,y\,\frac{d}{dy}\left[\frac{e^{-2yR}}{R^5}\left(4+R^2+4yR-yR^3
  \right)^2\right]=}
  \nonumber \\
  &=&-\frac{\Delta n^2}{4}\int_{\Delta
  n}^2\left(\frac{10}{R^6}+\frac{1}{R^4}+\frac{1}{8R^2}\right)\,dR
  \nonumber \\
  &=&\frac{1}{8}\left(\frac{\Delta n^2}{3}-\frac{4}{\Delta n^3}-
  \frac{2}{3\Delta n}
  -\frac{\Delta n}{4}\right)\,.\label{long}
\end{eqnarray}
We have put here $y_0=\infty$ without getting the divergencies. As
it is explained in the Appendix, in Eq.\ (\ref{long}) we have to
pick up only the term proportional to $\Delta n^2$.
Remarkably that this term is finite. It is an essential advantage
of our approach. The rest of the terms in this equation are
irrelevant to our consideration.
Thus the counter term for $E_W$ vanishes due to the regularizations
employed (see the Appendix).
{}In view of this we have
\begin{equation}
\label{EWF}
E_W^{\text{ren}}=
E_W=\frac{1}{2\pi a}\,\frac{1}{8}\,\frac{\Delta
  n^2}{3}=\frac{1}{48}\,\frac{\Delta n^2}{\pi a}\,.
\end{equation}
Finally we arrive at the following result for the Casimir energy
of a dilute dielectric ball
\begin{equation}\label{EFN}
  E^{\text{ren}}=E_W^{\text{ren}} +E_P^{\text{ren}}
=\frac{\Delta n^2}{\pi a}\left(\frac{1}{48}+
  \frac{5}{128}\right)=\frac{23}{384}\,\frac{\Delta n^2}{\pi a}\,.
\end{equation}
Taking into account the relation (\ref{epn}) between
$\varepsilon_i$ and $n_i$, $i=1,2$, we can write
\begin{equation}\label{F}
  E^{\text{ren}}=\frac{23}{1536}\,\frac{(\varepsilon_1-
 \varepsilon_2)^2}{\pi a}\,.
\end{equation}

     At the first time, this value for the Casimir energy of a dilute
dielectric ball has been derived in Ref. \cite{MNg2} by summing up the
van der Waals interactions between individual molecules making up the
ball ($\varepsilon_2 =1$). The
result (\ref{F}) was obtained also by treating a dilute dielectric
ball as a perturbation in the complete Hamiltonian of the
electromagnetic field for relevant configuration~\cite{Barton}. In
papers \cite{BMM,BM}, the value close to the exact one (\ref{F})
has been obtained by employing the uniform asymptotic expansion of
the Bessel functions.

     In Ref.~\cite{BNP} the estimation of the Casimir energy of a
dilute dielectric ball has been done taking into account, as it is
clear now, only the second term in Eq.\ (\ref{EFN}). And nevertheless
it was not so bad having the accuracy about $35\%$.

\section{Conclusion}
In this paper the exact (in the $\Delta n^2$--approximation) value of
the Casimir energy of a dilute dielectric ball is derived in the
framework of the quantum field theory. The starting point is the mode
summation by making use of the contour integration in the complex
frequency plane. Unlike the other approaches to this problem, we do
not use the uniform asymptotic expansion of the Bessel functions.

The key point in our consideration is employment of the addition
theorem for the Bessel functions which enables us to do the summation
over the angular momentum values in a closed form. As a by-product,
it is shown that the role of the contact terms, at least in the
$\Delta n^2$--approximation, consists only in removing the linear in
$\Delta n$ contributions to the Casimir energy. They do not
contribute to the finite value of this energy.

\acknowledgments
     The research has been supported by fund MURST ex 40\% and 60\%,
art.  65 D.P.R. 382/80.  The work was accomplished during the visit
of V.V.N. to the Salerno University. It is a pleasure for him to
thank Professor G.\ Scarpetta, Drs. G.\ Lambiase and A. \ Feoli for
warm hospitality. The financial support of IIASS and INFN is
acknowledged.  G.L. thanks the UE fellowship, P.O.M. 1994/1999, for
financial support.  V.V.N. is partially supported by Russian
Foundation for Basic Research (Grant No.\ 00-01-00300), he thanks
Professor G.\ Barton for providing an electronic copy of his recent
preprint \cite{Barton-dis}.

\appendix
\section*{Analysis of the divergencies generated by $W_{\lowercase{l}}^2$}
Here we reveal an important relation between linear and quadratic
in $\Delta n$ terms in $W_l^2$ defined in Eq.\  (\ref{W}).

Let us put
\begin{equation}\label{A1}
  x_1=y\left(1+\frac{\Delta n}{2}\right)\,,\quad
  x_2=y\left(1-\frac{\Delta n}{2}\right)\,,\quad
  \Delta x=\Delta n\,y\,.
\end{equation}
The Taylor expansion yields
\begin{eqnarray}
  W_l(x_1,
  x_2)&=&s_l(x_1)e_l^{\prime}(x_2)-s_l^{\prime}(x_1)e_l(x_2) \nonumber
  \\
  &=&-1+\left(2s_l^{\prime}e_l^{\prime}-s_le_l^{\prime\prime}-
             s_l^{\prime\prime}e_l\right)\,\frac{\Delta
             x}{2}\nonumber \\
 & &+\left[\frac{1}{2}\left(s_le_l^{\prime\prime\prime}-
        s_l^{\prime\prime\prime}e_l\right)
        +\frac{3}{2}\left(s_l^{\prime\prime}e_l^{\prime}-
         s_l^{\prime}e_l^{\prime\prime}
        \right)\right]\,\frac{\Delta x^2}{4}+O(\Delta x^3)\,. \label{A2}
\end{eqnarray}
For brevity we have dropped the argument $y$ of the function $s_l$ and
$e_l$, and have used the value of the Wronskian
\begin{equation}\label{A3}
  W\{s_l(y), e_l(y)\}=s_le_l^{\prime}-s_l^{\prime}e_l=-1\,.
\end{equation}
By making use of the equation for the Riccati--Bessel functions
\begin{equation}\label{A4}
  w_l^{\prime\prime}(y)-L(l, y)\,w_l(y)=0\,,\quad L(l, y)\equiv
  1+\frac{l(l+1)}{y^2}\,,
\end{equation}
we obtain
\begin{eqnarray}
 s_l^{\prime\prime\prime}e_l-s_le_l^{\prime\prime\prime}&=&L(l,
 y)\,, \nonumber \\
 s_l^{\prime\prime}e_l^{\prime}-s_l^{\prime}e_l^{\prime\prime}&=&-
            L(l, y)\,.  \label{A5}
\end{eqnarray}
Substitution of (\ref{A5}) into (\ref{A2}) gives
\begin{equation}\label{A6}
 W_l(x_1. x_2)=-1+[s_l^{\prime}e_l^{\prime}-L(l, y)s_le_l]\Delta
 x-\frac{1}{2}\,L(l, y)\,\Delta x^2 +O(\Delta x^3)\,.
\end{equation}
Squaring Eq.\  (\ref{A6}) one gets
\begin{equation}\label{A7}
  W_l^2(x_1, x_2)=1+A_l\,\Delta n+B_l\,\Delta n^2+O(\Delta n^3)\,,
\end{equation}
where
\begin{eqnarray}
  A_l&=&y(s_l''e_l+s_le_l''-2s_l'e_l')=2y\left[2L(l,
  y)s_le_l-\frac{1}{2}\,(s_le_l)^{\prime\prime}\right]\,, \label{A8}\\
  B_l&=&y^2L(l, y)+\frac{1}{4}\,A_l^2\,. \label{A9}
\end{eqnarray}

In terms of these notations we can write
\begin{equation}\label{A10}
  \ln \left(W_l^2-\frac{\Delta n^2}{4}\,P_l^2\right)=
  A_l\Delta n+\left(B_l-\frac{A_l^2}{2}\right)\Delta n^2-
  \frac{\Delta n^2}{4}\,P_l^2+O(\Delta n^3)\,.
\end{equation}
The terms quadratic in $\Delta n$ in Eq.\  (\ref{A10}) exactly
reproduce the function $F_l(y)$ in Eq.\  (9) of the paper \cite{BMM}.
It is this function that affords the whole finite value of the
Casimir energy in the problem under consideration.  Unlike the papers
\cite{BMM,BM,Mar} we didn't introduce the contact terms in the
definition of the Casimir energy and nevertheless we have reproduced
the key function $F_l(y)$. It implies that the contact terms do not
give a contribution into the finite part of the Casimir energy in the
problem under consideration.  They  merely cancel the terms
$A_l\Delta n$ in Eq.\ (\ref{A10}).

Now we show, without invoking the contact terms,
that the $A_l$ terms in Eq.\ (\ref{A10})
do not contribute into the vacuum energy.

Using  Eq.\ (2.19) with $\theta =0$ we introduce the notation
\begin{equation}
\sum_{l=1}^\infty (2l+1)s_l(yr)e_l(y\rho)+1 =\frac{yr\rho}{|r-\rho|}
e^{-y|r-\rho|}\equiv \overline {\cal D}(r,\rho,y)\,{.}
\label{A11}
\end{equation}
Taking into account the explicit form of the coefficients $A_l$
defined in Eq.\ (\ref{A8}) one can write
\begin{equation}
\Delta n\sum_{l=1}^\infty (2l+1)A_l=\left . y\Delta n \left(
\frac{\partial ^2}{\partial r^2}-2\frac{\partial ^2}{\partial r
\partial \rho}
+\frac{\partial^2}{\partial \rho^2}
\right )
\overline{\cal D}(r,\rho,y)\right |_{r=\rho=1}+1\,{.}
\label{A12}
\end{equation}
When $r=\rho=1$ the derivatives of the function ${\cal \overline D}$
in Eq.\ (\ref{A12}) tend  to infinity. Therefore a preliminary
regularization should be introduced here in order to put our
consideration on a rigorous mathematical footing. To this end we
define the right-hand side of Eq.\ (\ref{A12}) in the following way
\begin{equation}
\Delta n\sum_{l=1}^\infty (2l+1)A_l=\left .\Delta n
 \lim_{\varepsilon \to 0}
\left( \overline{\cal D}_{rr} -2\overline{\cal D}_{r\rho}
+\overline{\cal D}_{\rho \rho}
\right)\right |_{r=1+{\varepsilon}/{2}
\atop \rho=1-{\varepsilon}/{2}} +1\,{,}
\label{A13}
\end{equation}
where the positive constant $\varepsilon $ is a regularization
parameter. From the explicit form of the function $\overline {\cal
D}(r,\rho,y)$ (see Eq.\ (\ref{A11})) it follows immediately
\begin{equation}
\left .\lim_{\varepsilon \to 0}
\left( \overline{\cal D}_{rr} -\overline{\cal D}_{\rho \rho}
\right)\right |_{r=1+{\varepsilon}/{2}
\atop \rho=1-{\varepsilon}/{2}} =0\,{.}
\label{A14}
\end{equation}
The analogous limit for the differences
\begin{equation}
 \overline{\cal D}_{rr} -\overline{\cal D}_{r \rho} \quad \text{ and }
 \quad
 \overline{\cal D}_{\rho \rho} -\overline{\cal D}_{r\rho}
\label{A15}
\end{equation}
also vanishes. Hence  in the regularization introduced above the sum
under consideration has the following value
\begin{equation}
\Delta n\sum_{l=1}^\infty (2l+1)A_l=1\,{.}
\label{A16}
\end{equation}
It implies immediately that the term linear in $\Delta n$, which
encounters Eq.\ (\ref{A10}) does not contribute into the vacuum
energy $E_2$ defined in Eq.\ (\ref{E2}).

  Now we show that the contributions into the Casimir energy given by
$\sum_lB_l$ and by $(1/4)\sum_lA_l^2$ are the same. In other words,
$y^2L(l, y)$ in Eq.\  (\ref{A9}) does not give any finite
contribution into the vacuum energy. In order to prove this, we
consider the expression
\begin{equation}\label{A17}
  I=\sum_{l=1}^{\infty}\nu\int_0^{\infty}y^2\,dy\,,\quad
  \nu=l+\frac{1}{2}\,{.}
\end{equation}
Instead of the cutoff regularization we shall use here the analytical
regularization presenting (\ref{A17}) in the following form
\begin{eqnarray}
  I &=& \lim_{s\to 0}\sum_{l=1}^{\infty}\nu\int_0^{\infty}y^{2-s}dy=
       \lim_{s\to 0}\sum_{l=1}^{\infty}\nu^{4-s}
       \int_0^{\infty}z^{2-s}dz\nonumber \\
    &=& \lim_{s\to 0}\lim_{\mu^2\to 0}\sum_{l=1}^{\infty}\nu^{4-s}
       \int_0^{\infty}(z^2+\mu^2)^{1-s/2}dz\,. \label{A18}
\end{eqnarray}
Here the change of integration variable $y=\nu z$ is done and the
photon mass $\mu$ is introduced. Further we have
\begin{eqnarray}
 I&=&\lim_{s\to 0}\lim_{\mu^2\to 0}[(2^{-4+s}-1)\zeta
 (s-4)-2^{-4+s}]\,\frac{\mu^{3-s}}{2}{
 \frac{\Gamma {\left(\displaystyle\frac{1}{2}\right)}
 \Gamma\left(\displaystyle
  {-\frac{3}{2}+\frac{s}{2}}\right)}{\Gamma\left(
 \displaystyle\frac{s}{2}-1\right)}} \nonumber
 \\
 &=& -\frac{\pi}{24}\lim_{s\to 0}\lim_{\mu^2\to 0}
 \frac{\mu^2}{\Gamma\left (\displaystyle\frac{s}{2}-1
\right )}\to 0\,. \label{A19}
\end{eqnarray}

In view of all this, we are left with the following scheme for
calculating the Casimir energy in the $\Delta n^2$--approximation in
the problem under consideration.  First, the $\Delta
n^2$--contribution should be find, which is given by the sum
$\sum_lW_l^2$. Upon changing its sign to the opposite one, we obtain
the contribution generated by $W_l^2$, when this function is in the
argument of the logarithm. Obviously, this result would be deduced
directly if one could find in a closed form the sum
$\sum_lW_l^2W_l^2$ \cite{Klich}.  This assertion can be explained  by
a symbolic formula
\begin{equation} \ln \left(W_l^2-\frac{\Delta
n^2}{4}\,P_l^2\right) \sim -\Delta n^2 B_l- \frac{\Delta
n^2}{4}\,P_l^2+O(\Delta n^3)\,{.} \label{A20}
\end{equation}
The sign $\sim $ means here the equality subject to the
regularizations described above are employed.

\end{document}